\pdfoutput=1

\documentclass[11pt]{article}

\usepackage[]{ACL2023}

\usepackage{times}
\usepackage{latexsym}

\usepackage[T1]{fontenc}
\usepackage[utf8]{inputenc}
\usepackage{csquotes}

\usepackage{amsmath,amssymb}
\usepackage{subcaption} 

\usepackage{tcolorbox}
\usepackage{makecell}
\usepackage{algorithmic}
\usepackage[linesnumbered,ruled,vlined]{algorithm2e}
\usepackage{geometry}
\usepackage{quoting}

\usepackage{graphicx}
\usepackage{xcolor} 
\usepackage{caption}

\usepackage{amssymb}
\usepackage{multirow}
\usepackage{array}
\usepackage{amsmath}
\usepackage{booktabs}
\usepackage{makecell}
\usepackage{url}
\usepackage{float}
\usepackage{colortbl}
\usepackage{array}
\usepackage{booktabs}
\usepackage{tabularx}
\usepackage{multirow}
\usepackage{listings}

\usepackage[utf8]{inputenc}

\usepackage{microtype}

\usepackage{inconsolata}

\title{Attention Mechanism for LLM-based Agents \\ Dynamic Diffusion under Information Asymmetry}

\author{Yiwen Zhang$^1$, Yifu Wu$^2$, Wenyue Hua$^3$, Xiang Lu$^2$, Xuming Hu$^1$ \\
$^1$The Hong Kong University of Science and Technology (Guangzhou) \\
$^2$Independent Researcher 
$^3$University of California, Santa Barbara \\
}

\begin{document}

\maketitle

\begin{abstract}
Large language models have been used to simulate human society using multi-agent systems.
Most current social simulation research emphasizes interactive behaviors in fixed environments, ignoring information opacity, relationship variability, and diffusion diversity.
In this paper, we first propose a general framework for exploring multi-agent information diffusion. 
We identified LLMs' deficiency in the perception and utilization of social relationships, as well as diverse actions.
Then, we designed a dynamic attention mechanism to help agents allocate attention to different information, addressing the limitations of the LLM attention mechanism.
Agents start by responding to external information stimuli within a five-agent group, increasing group size and forming information circles while developing relationships and sharing information. 
Additionally, we explore the information diffusion features in the asymmetric open environment by observing the evolution of information gaps, diffusion patterns, and the accumulation of social capital, which are closely linked to psychological, sociological, and communication theories.

\end{abstract}

    \begin{quoting}
        \small
        ``\textit{The truth is rarely pure and never simple.}'' \begin{flushright} ------ \textit{The Importance of Being Earnest} \end{flushright}
    \end{quoting}


\section{Introduction}
Recent advances in large language models (LLMs) with strong reasoning and language understanding ability have established a robust foundation for developing agents that exhibit social intelligence \cite{b1}. Many studies have employed LLM-based agents to simulate human behavior, construct social networks, and explore various dimensions of social development and human conduct \cite{m56,b3}. For instance, researchers have investigated the social capabilities of these agents by modeling market competition \cite{m37}, economic flows \cite{m59}, international trade \cite{b2}, warfare \cite{m60}, and political party competition \cite{m63}, thereby providing insights and recommendations for real-world applications. However, these simulations often operate within fixed environments \cite{m9} or assume static channels for information transmission \cite{b55}. As a result, they often overlook the role of information opacity, \emph{i.e.}, the asymmetric distribution of information, which can profoundly influence actual human decision-making processes and, consequently, the validity of the simulation outcomes.

Real-world information is neither transparently nor equally distributed, leading to inherent information asymmetry \cite{b13}. Typically, individuals acquire information in a progressive, staged, and selective manner \cite{b37,b38}, with the effectiveness of this process depending on both the methods employed and the individual’s interpretive abilities. Consequently, organizations such as businesses \cite{b5}, prosecution agencies \cite{b10}, government systems \cite{b6,b8}, news media \cite{b9}, and software developers \cite{b7} have developed strategies to tailor the disclosure of information, thereby facilitating easier access. Moreover, interpersonal communication and the formation of social connections further enable individuals to obtain additional details \cite{b14}. Given the diversity of social networks, the nature and extent of the information that individuals receive are significantly shaped by their social interactions.

In this project, we investigate the dynamics of information diffusion within an asymmetric open environment using a multi-agent simulation framework. An information asymmetry situation refers to a scenario where one party in a transaction or interaction possesses more, or higher quality, information than the other potentially due to varied information sources, evolving relationships, and differing contents of information. By comparing simulation outcomes with predictions derived from real-world information theory \cite{b62,b64}, we aim to understand how agents cope with asymmetric information and whether their behaviors mirror those of humans. We hope to enhance the validity of multi-agent social simulations under conditions of information asymmetry and to demonstrate that LLM-based agents can effectively simulate complex social dynamics.

To achieve this objective, we first introduce a two-tier general simulation framework designed to capture dynamic information diffusion. We also propose an agent attention mechanism \cite{b51,b49} that prioritizes critical information in a manner analogous to human information processing, enabling agents to handle multiple sources of information concurrently. We then examine the behaviors of agents under various external stimuli. 
The Dynamic Attention algorithm helps to reduce the Action Similarity Bias and increase the Relationship Perception Frequency, thereby proving the effectiveness of the algorithm in assisting agents in handling complex social scenarios in social simulations. Meanwhile, we explored the social phenomenon of multi-agent information diffusion in the asymmetric open environment.




\definecolor{bluee}{RGB}{178, 139, 103}
\newcommand{\GetMaxAgent}[1]{...}

\section{Method}



\subsection{General Simulation Framework}

The simulation framework consists of two stages: the initial stage and the interaction stage. The initial stage is the pre-simulation setup, which includes selecting groups characterized by specific topological structures from various social networks and defining their corresponding profiles and relationships. The interaction stage encompasses the entire process of agent interaction during the simulation.

The initial stage establishes the foundational social network. Drawing upon principles of organizational behavior in social science \cite{b50,b46}, we select two representative network topologies: the wheel and the circle \cite{b46}. The wheel structure is characterized by a central node connected to multiple peripheral nodes, forming a centralized network, whereas the circle structure involves peripheral nodes interconnected in a circular manner, representing a decentralized network. The network comprises five agents, which is the minimum number necessary to distinguish between these two topological configurations. These agents are allowed to disclose only their profiles to the external environment, while their subjective relationships, actions, and memories remain private.

During the interaction phase, the simulation is conducted over ten rounds, during which all agents can send messages to any other agent within both the initial setup and the open environment. In this context, the term ``open environment'' refers to the allowance for an indefinite number of new agents with diverse profiles. For instance, if an agent wishes to communicate with a police officer and no such agent currently exists in the environment, the agent may define a new profile and relationship for a police officer and incorporate this new agent into the current group. This mechanism is designed to emulate an open environment where any type of agent can be encountered. In each round, agents have the flexibility to either disseminate information or modify their relationships. The simulation framework's support for an unbounded network size enables agents to distribute information without limitation.

\begin{figure}[!ht]
    \centering
    \includegraphics[width=\linewidth]{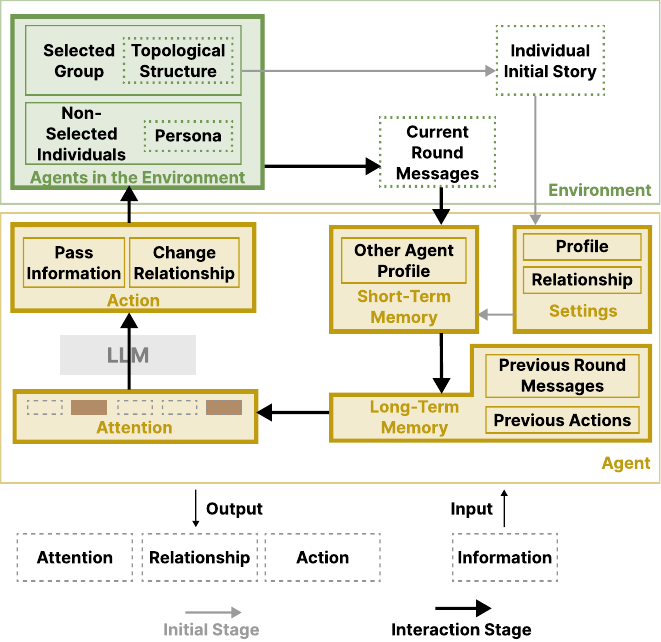}
    \caption{The two-stage framework model to simulate asymmetric open environment information diffusion.}
    \label{fig:Simulation}
\end{figure}

\begin{table*}[!ht]
    \centering
    \begin{tabular}{@{}*{9}{l}@{}} 
        \toprule
         &  \multicolumn{4}{c}{Generic LLM Agent} &  \multicolumn{4}{c}{Dynamic Attention Agent} \\ 
         \cmidrule(lr){2-5} \cmidrule(lr){6-9} 
        & \multicolumn{2}{c}{SH-BC-negative} & \multicolumn{2}{c}{OG-BCR-circle} & \multicolumn{2}{c}{SH-BC-negative} & \multicolumn{2}{c}{OG-BCR-circle} \\ 
        \cmidrule(lr){2-3} \cmidrule(lr){4-5} \cmidrule(lr){6-7} \cmidrule(lr){8-9} 
        & circle & wheel & positive & negative & circle & wheel & positive & negative \\ 
        \midrule
        \multicolumn{9}{c}{Action Similarity Bias $\downarrow$} \\ 
        \midrule
        \texttt{GPT-4o-mini} & 0.99 & 0.99 & 0.95 & 0.92 & 0.90 & 0.90 & 0.86 & 0.87 \\ 
        \texttt{llama-3.3-70b} & 0.96 & 0.96 & 0.94 & 0.94 & 0.89 & 0.89 & 0.87 & 0.88 \\ 
        \midrule
        \multicolumn{9}{c}{Relationship Perception Frequency $\uparrow$} \\ 
        \midrule
        \texttt{GPT-4o-mini} &0.67 & 0& 0.67& 2& 2&2.67 & 2& 5.33\\ 
        \texttt{llama-3.3-70b} & 0&0 &0 & 0& 0.33& 0.33& 0& 0.33 \\ 
        \bottomrule
    \end{tabular}
    \caption{Preliminary test for LLM's deficiency on information diffusion in social simulations. The table shows the mean value for three times simulation for every setting.}
    \label{tab:finding12}
\end{table*}




\noindent \textbf{Action} \quad At each time step $ T = \left \{ t_1,...,t_{10} \right \} $, we have agents $ A = \left \{ a_1,...,a_n \right \} $. At the beginning of the simulation time step $t_1$, $n=5$. At each time step $t$ after this, n may increase based on the actions of each agent, up to a maximum of 5 per round.
Each agent $a_i$ has profile $p_i$, relationship $r_i$, output action $o_i$, information diffusion $d_i$.
At time step $t_i$, $p_i$ remains unchanged, $r_i$ has scale $r_i \in \left \{ \text{positive}, \text{negative}, \text{general} \right \}$ with other agents.
$o_i$ can be True or False and consists of two parts: changing the relationship $r_i$ and transmitting information $d_i$.
The agent can independently choose to pass information to any agent in the current environment or to a new agent it defines itself.
Therefore, the agent's action decision-making must balance the initial information with other information, including discussions caused by profile similarities.
After each action round, the environment updates the state of each agent based on the agent's actions $ O = \left \{ o_1,..., o_n \right \} $. This includes updating $r_i$ (subjective relationship) in the database, adjusting $d_i$ to reflect the corresponding receiver's received\_messages, and refreshing the agent's actions for this round.
After that, the environment updates the list of the latest agents and performs attention calculations as algorithm \ref{alg1} and action decisions for the next round of agents. More detailed explanations are available in Appendix \ref{appb}.

\subsection{LLM's Deficiency in Complex Social Environments}

After building the general simulation framework, we set up preliminary tests for learning generic LLM competency. We employed a generic LLM-based agent that retained all information received across multiple rounds in its memory, making decisions based on this complete dataset. We find two kinds of deficiency of generic LLM on information diffusion in social simulations. The test experiment settings include four kinds of information asymmetry environments, and the detailed information list in Appendix \ref{appendix env} (all the simulations ran by SOTA models \texttt{GPT-4o-mini-2024-07-18} and \texttt{llama-3.3-70b-instruct}).

\paragraph{\textit{LLM outputs similar actions in different rounds.}}


We observed that, under this design, a single agent's actions across communication rounds remained highly similar throughout the simulation. This indicates that agents relying on generic LLMs have difficulty generating diverse and meaningful information dissemination behaviors. Such uniformity in behavior diverges from patterns typically observed in human interactions \cite{b73,b74,b75}. In contrast, Dynamic Attention helps the agent reduce the likelihood of outputting similar messages in multiple rounds, thereby increasing the diversity of the messages.

\paragraph{\textit{LLM lacks perception of social relationships.}}

Another finding focuses on how the LLM agent uses the relationship and how the relationship changes during the simulation.
We observe LLM remains a low relationship perception frequency in different asymmetry environments, which means the social relationship cannot effectively influence LLM's decision and is also not included in the decisions.
This observation stands in stark contrast to real-world information diffusion processes \cite{b76,b77}. Consequently, relying solely on an LLM’s intrinsic attention mechanisms over an extended context constrains the representation of how various pieces of information (sent by other agents with different social relationships) compete for an agent’s focus. 
Agents need more factors related to the real world (such as interpersonal relationships, information complexity, and information changes) to assist them in making wise action decisions.
To address these shortcomings, we propose an agent attention algorithm designed to mitigate these issues.


\subsection{Agent with Dynamic Attention}

The Dynamic Attention Mechanism is grounded in research from social science and journalism, particularly the idea that multiple pieces of information compete for an individual’s attention, as articulated by the Global Workspace Theory \cite{b51}. In the context of transformer-based models, biases introduced during pre-training \cite{b67} and the ``lost in the middle'' issue associated with lengthy text inputs \cite{b66} underscore the need for an algorithmic approach that enables agents to dynamically prioritize crucial information. Accordingly, agents must adapt their focus to evolving inputs and thoroughly evaluate the importance of new data before deciding on a course of action. Insights from journalism further guide this design: people’s attention is often heightened by enhancing the relevance of the information, citing significant sources, and foregrounding key points \cite{b68}. Building on these principles, our mechanism determines whether the agent should prioritize certain pieces of information and adjusts the presentation of historical messages to better reflect their relative importance. Below is the algorithm:

\begin{algorithm}[ht]
\small
\caption{Dynamic Attention Algorithm}
\label{alg1}
\KwIn{received\_messages, turn\_number, actions, subjective\_relationships}
\KwOut{attention\_information}

$\text{current msgs}, \text{prev msgs} \gets \{(s, m) | (t, s, m) \in \text{received messages}, t = \text{turn number}\}$\;

\ForEach{$(s,m) \in \text{msgs}$}{
    $r \gets rel.get(s)$\;
    $w \gets \begin{cases}
        1, & \text{if } r \in \{\text{pos,neg}\}\\
        0, & \text{if } r = \text{gen}\\
        -1, & \text{otherwise}
    \end{cases}$\;
    $dict[s] \gets (w,m)$\;
}

$\text{max\_agent} \gets \text{GetMaxAgent}(\text{CalcEntropy}(\text{msgs}))$\;

\If{$\text{max\_agent} \neq \emptyset$}{
    \ForEach{$(s, \text{info}) \in weight\_dict$}{
        $\text{info[weight]} \gets \text{info[weight]} + \begin{cases}
            1, & \text{if } s = \text{max\_agent}\\
            -1, & \text{otherwise}
        \end{cases}$\;
    }
}

\ForEach{$(s, \text{info}) \in weight\_dict$}{
    \If{$s \in prev\_dict$}{
        $\text{prev\_entropy} \gets \text{CalcEntropy}(prev\_dict[s])$\;
        $\text{curr\_entropy} \gets \text{CalcEntropy}(prev\_dict[s] \cup \{\text{info[message]}\})$\;
        $\text{info[weight]} \gets \text{info[weight]} + \begin{cases}
            1, & \text{if } \text{curr\_entropy} > \text{prev\_entropy}\\
            -1, & \text{otherwise}
        \end{cases}$\;
    }
}

\If{$\text{actions} \neq \emptyset$}{
    $top\_agent \gets \GetMaxAgent(\text{Counter}(actions))$\;
    \ForEach{$(s, \text{info}) \in weight\_dict$}{
        $\text{info[weight]} \gets \text{info[weight]} + \begin{cases}
            1, & \text{if } s = \text{top\_agent}\\
            -1, & \text{otherwise}
        \end{cases}$\;
    }
}
\end{algorithm}

Algorithm \ref{alg1} outlines the procedure through which an agent processes multiple pieces of incoming information to compute the importance weight of each message, leveraging both short-term and long-term memory. The algorithm takes as input the agent’s previously received messages, past actions, most recent subjective relationships, and the current simulation round number. Its output is a weighted information set for all messages received in the present round.

Initially, the algorithm distinguishes between newly received messages and those stored from previous rounds. The short-term memory component only includes messages from the current round and the most recent subjective relationships, while the long-term memory component holds all previous messages and actions. The weighting process begins with an initial assessment in short-term memory, simulating the quick human evaluation of multiple messages over a brief time span. First, the relationship between the message sender and the agent is determined: agents with a positive or negative relationship receive an increased weight, while neutral relationships remain unaltered, and unfamiliar agents lead to a reduced weight. Among all messages received in the current round, those deemed “high complexity” also receive higher weights due to their novel information content. This preliminary weighting is performed at a relatively low computational cost.

Subsequently, the agent refines these weights by comparing short-term memory with long-term memory. This step emulates the process by which humans recall information sources and consider past exchanges. To highlight messages that exhibit the greatest level of transformation during transmission, the algorithm calculates the change in the entropy value of the corresponding information source from the previous round to the current round. For each round, if the message output by the agent is $p$, then calculate the entropy value of this agent in round $i$ according to Equation \ref{shanon}. Lastly, the algorithm further increases the weight of messages originating from agents with whom there have been the most frequent interactions, as inferred from past actions.

\begin{equation}
    H(X) = -p_i \log (p_i)
\label{shanon}
\end{equation}

\section{Experiment and Evaluation}

This section outlines our experiment design and the ablation study for Dynamic Attention Algorithm. We establish 12 types of asymmetric environments based on information diffusion and test the performance of the SOTA LLM using the Dynamic Attention algorithm. In the ablation experiment, we compare with generic llm (with all memory) and generic llm (with last round memory), and verified the effectiveness of the algorithm from two perspectives: the agents' behavior of and the utilization of social relationships.

\subsection{Experimental Settings}
\label{3.1}

To examine the formation of agent information diffusion in an asymmetric open environment, we conduct a simulation and tested it in 12 different asymmetrical information environments.
The simulation is developed based on the SOTOPIA \cite{b53} library, Redis database \cite{b80}, and employs the SOTA model \texttt{GPT-4o-mini} and \texttt{llama-3.3-70b-instruct} \cite{b54,b78} for the agent's decision-making process.
We randomly select 5 agents from the 25 agents in Stanford Town \cite{m9} as the initial state group. 
Their profiles include gender, age, innateness, and occupation, and are evenly distributed. 
The group settings include the group's topology and initial relationship.

We jointly build an information asymmetry environment through \textit{information content} and \textit{distribution mechanism}. 
The main difference in the information content lies in its relevance to initial agents, and the distribution mechanism mainly affects the asymmetry generated directly at the source of information.
Based on the \textbf{Construal Level Theory} \cite{b69} in social psychology, we define four types of information content: \textbf{other people's gossip (OG), public policy (PP), legal cases (LC), and stakeholder information (SH).} 
Furthermore, we define three distribution mechanisms: \textbf{information broadcast (BC), information unicast (OA), and broadcast by round (BCR),} creating asymmetry at the source of information.
BC represents the process of send the information to all five agents at the first round, while OA means only send information to one agent (agent 2 as the center in wheel and common node in circle).
BCR means send information to one agent each round until the initial five agents know the information.
Our information is generated by \texttt{GPT-4o-mini} and is about 50 words long, as shown in table \ref{tab:content_analysis}.

We run simulations three times for each topology corresponding to the information content and asymmetric mechanism, with the initial relationships between agents set to all positive or all negative.

\subsection{Ablation Study} 

In this section, we compare three scenarios: the dynamic attention algorithm (Dynamic Attention), the generic LLM with all-round memory (Generic LLM-all), and the generic LLM with last-round memory (Generic LLM-last). Based on the experiment settings, we choose different asymmetry environments to verify the dynamic attention algorithm. We list the results in Appendix \ref{appendix ablations}.

\newtcolorbox{mybox}{colframe = green!25!black}


\section{Result and Analysis}

\textbf{RQ 1: How do asymmetry settings influence agents' information diffusion? }

\begin{mybox}
Distributing information over time helps maintain relevant knowledge within a group, but it is not effective for widespread sharing. 
\end{mybox}

\begin{figure*}[!ht]
    \centering
    \includegraphics[width=\linewidth]{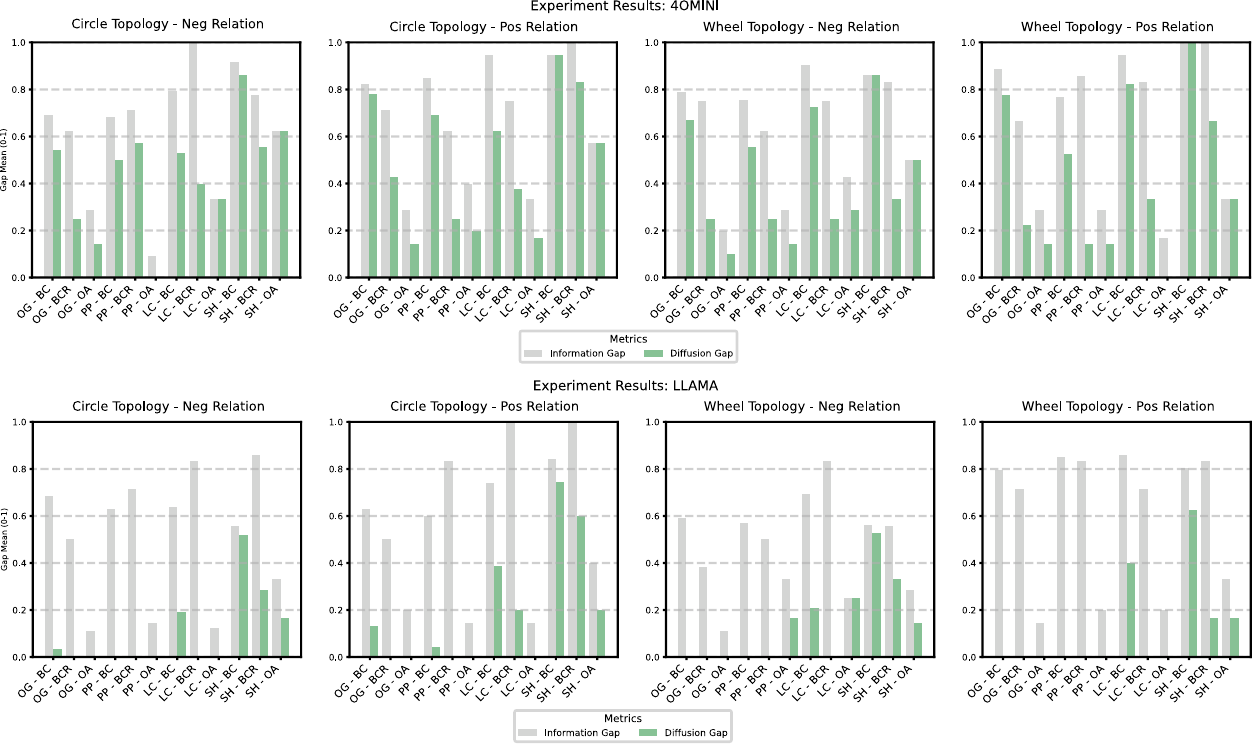}
    \caption{Information Gap (the grey bars) and Diffusion Gap (the green bars) for 12 asymmetric environments on four initial settings. Each simulation contains these two values. Differences between the two values represent the Diffusion Conversion Gap. The smaller the gap, the more individuals with known information tend to spread it, which means that the diffusion chain is relatively complete.}
    \label{fig:gap}
\end{figure*}

We calculate the Information Retention for every simulation, and the result shows simulations using the BCR mechanism remain a 3.14 rounds for the initial information retention in the group diffusion process, while in the BC setting, the value is only 2.77.
As shown in Figure \ref{fig:gap}, in different models, we can observe that although the BCR mechanism keeps the initial information in the population for a longer period compared to the BC mechanism, it does not lead to more spread within the population. This phenomenon is in line with the \textbf{Agenda-Setting Theory} in communication studies. When the source of information uses different methods to disseminate the initial information, the group will exhibit corresponding different behaviors.

\begin{mybox}
    Different LLMs may vary in their willingness to spread information, but their performance trends are consistent in different environments of information asymmetry.
\end{mybox}

We observe similar trends in the differences among the three information distribution mechanisms. This indicates that the information gap and diffusion gap reach their highest levels when the information source broadcasts. Following this is BCR, and then comes OA. 
We can also observe that the \texttt{llama-3.3-70b-instruct} model consistently shows differences in its willingness to disseminate information content across various initial group relationships. 

\begin{mybox}
    The agent tends to disseminate internally the information that they are the party involved, while spreading externally information that is beneficial to the public.
\end{mybox}

Although the gap in the diffusion intentions presented by the \texttt{gpt-4o-mini} model is relatively small, it can still be observed that both models are more inclined to disseminate information that is related to their interests or where they are the parties involved. 
More importantly, based on the values of LC and SH in Table \ref{tab:newagentana}, when the agent is the party involved in the event, they tend to have internal discussions rather than disseminate the information to other agents outside. The difference is that the SH information not only includes the initial five aspects of the agent's interests, but also contains beneficial information for the public. In such cases, the agent is more likely to disseminate this information to the outside world.

\noindent \textbf{RQ 2: How does social relationship and topology change after the agent's information diffusion in "open environment"? }

In our design, the agent can choose to send messages to any new agent (which does not exist in the environment) at any round. But this does not mean that they will choose to pass on the relevant content of the initial information to the new agent. Therefore, we counted the number of new agents in different information environments, as well as the similarity between the information received by the new agents upon their initial entry into the group and the initial information (shown in Table \ref{tab:newagentana}). 
In the stakeholder information environment, agents are most inclined to communicate with new agents, which is the opposite of the legal case scenario. Similarly, in the stakeholder environment, the content exchanged between the agent and the new agent is the most similar to the initial information..
Furthermore, all the cases where the new agent received information that was 80\% similar to the initial information were concentrated in the Stakeholder information.

\begin{table}[h]
    \centering
    \small
    \begin{tabular}{ccccc}\toprule
         &  OG&  PP&  LC& SH\\\midrule
         \makecell[c]{the average number\\for agent increase}&  1.196&  1.167&  0.944&1.375\\
         \midrule
         \makecell[c]{new agent \& receive\\ initial information}&  0.293&  0.267&  0.304&0.404\\ \bottomrule
    \end{tabular}
    \caption{Analysis for new agents in the asymmetric open environment}
    \label{tab:newagentana}
\end{table}

In the same experimental simulation setting, different models perform variably, which contrasts with common perceptions. We calculate the changes made by the model to the social relationship and topology (represented by the agent number). In Table \ref{tab:RQ2}, the ratio of actions that changed the existing relationships or added new agents to the total number of simulations is shown for all simulations under the corresponding environmental settings.
\texttt{gpt-4o-mini} tends to directly pull new agents into the group, while \texttt{llama-3.3-70b-instruct} usually makes changes within the existing relationships.

\begin{table}[h]
    \centering
    \small
    \begin{tabular}{ccccccl}\toprule
         &  \multicolumn{3}{c}{\texttt{gpt-4o-mini}}& \multicolumn{3}{c}{\texttt{\makecell[c]{llama-3.3-\\70b-instruct}}}\\\midrule
 & BC& OA& BCR& BC& OA&BCR\\
         \midrule
         \makecell[c]{change\\relationship}&  /&  /&  0.08&  1& 1.08&1.17\\
         \midrule
         \makecell[c]{add new\\agent}&  1.42&  1.25&  1.33&  0.08& /&0.17\\ \bottomrule
    \end{tabular}
    \caption{Comparison for different LLM social behavior in "asymmetric open environment".}
    \label{tab:RQ2}
\end{table}

\begin{figure*}[ht]
    \centering
    \includegraphics[width=\linewidth]{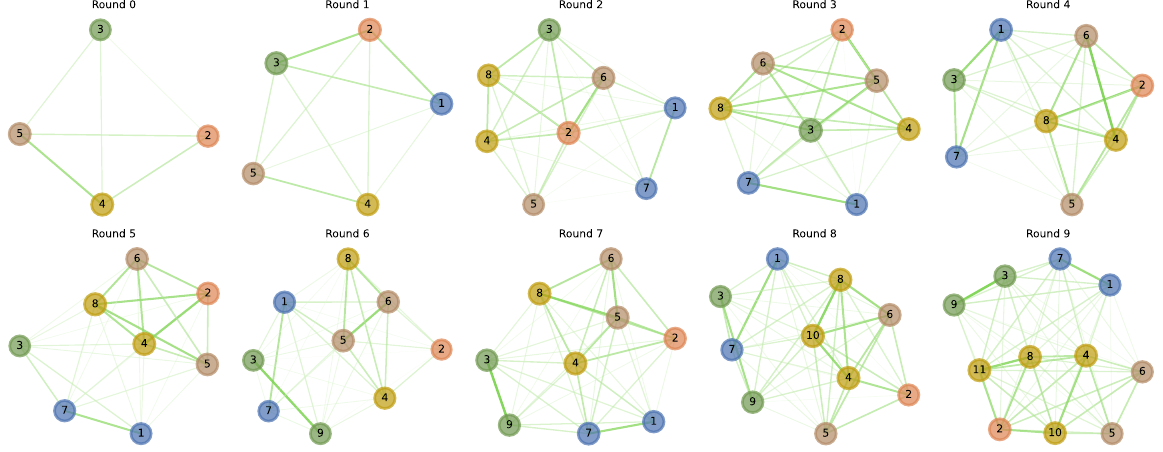}
    \caption{In this Social Capital Theory case study, agents 1, 3, 4, and 5 form relationships with new agents, creating distinct information circles within the growing group. Nodes represent individual agents, with colors indicating their lineage (the agent and its recruits are in the same lineage). Node distance and edge color depth reflect the similarity between the current message and those of all agents. Agents who did not take action are not recorded.}
    \label{fig:topologycase}
\end{figure*}

\begin{table*}[ht]
    \centering
    \small
    
    \scalebox{0.8}{
    \begin{tabular}{p{3cm}p{16cm}}
    \toprule
        cooperation & "Hey there! As a bartender and bar owner, ..., collaborate and share our ideas on making events special and enjoyable." \\
        \midrule
        \textcolor{bluee}{cooperation} & \textcolor{bluee}{"Hey! I came across a funding initiative that offers up to \$50,000 for innovative projects, ..., Let's explore our ideas together."} \\
        \midrule
        further discussion & "I'd love to discuss how we can challenge our preconceptions together and explore themes that resonate in both fields." \\
        \midrule
        \textcolor{bluee}{further discussion} & \textcolor{bluee}{"I've been reflecting on the recent events surrounding the disappearance of digital artwork, ..., What do you think this might symbolize in the context of our relationship with technology and existence?"} \\
        \midrule
        \textcolor{bluee}{altruism} & \textcolor{bluee}{"I came across an exciting funding initiative that offers up to \$50,000 for projects that, ..., this could really enhance your creative projects! Applications open next month, and I believe ..."} \\
        \midrule
        active inquiry & "I’d love to hear your thoughts on how we can use technology to enhance artistic expression." \\
        \midrule
        support & "I just want to share how much I appreciate your efforts in, ..., feel free to reach out to The Rose and Crown Pub. I'd love to support your initiatives!" \\
        \bottomrule
    \end{tabular}
    }
    \caption{Cases of agent's purpose for information diffusion. The diffusion of content can be related (colored red) or unrelated to the initial information and can serve a similar purpose.}
    \label{tab:behavior}
\end{table*}

\noindent \textbf{RQ 3: What's the relation between received messages and the agent's social behavior?}

In the experiments, we observe the social characteristics displayed by the agent when conveying information. For example, after discussing activity details, the agent may express a desire to cooperate or exhibit altruistic behavior by bringing other relevant agents into the group and relaying pertinent information. Moreover, regardless of whether the information sent by the agent is related to the initial information, it will always reflect social features. Specific instances are presented in Table \ref{tab:behavior}.

On the other side, different weights are assigned to the received messages to influence the agent's behavior by competing for its attention. The experimental results show that obtaining a high-weight message does not necessarily directly prompt the agent to spread this message, nor does it directly lead the agent to converse with the sender of the high-weight message in the next round. Due to the agent's influence or other factors, messages related to the initial information may not receive a high weight. This also explains why initial environmental information might cease to be disseminated in a certain round. This setup integrates the changes in social relationships and the messages themselves over multiple rounds into the agent's information diffusion process, guiding the LLM to take actions that better suit the agent's current situation. Detailed cases are provided in Appendix \ref{appendix:caseDA}.

Overall, we observe many cases of \textbf{Social Capital Theory} in different simulations. Although in our research, agents don't have economic settings, they also accumulate information capital by sending messages to other agents, pulling different new agents into the group, and building wise social relationships with initial and subsequent agents.

\section{Related Work}

\subsection{Information Asymmetry and Diffusion}

Information asymmetry \cite{b58,b35,b29} refers to the difference in information among parties in a transaction or interaction, where one party has more or better information than the other. 
There are two types of information asymmetry: asymmetric information, where one party is known but the other is not, and symmetric lack of information, where all parties are unknown \cite{b30}. 
Common information diffusion models, like the IC model \cite{b57} and SIR model \cite{b56}, use probabilistic approaches to simulate diffusion. 
While these models offer a structured framework, their reliance on mathematical constraints—such as individual activation probabilities and discrete states—limits their real-world applicability \cite{b55}.
In this paper, we use a simulation approach with LLM-based agents to explore complex social scenarios involving information asymmetry. 
By comparing our results with existing theoretical frameworks, we show that LLM-based agents exhibit behaviors similar to human information processing, validating the use of multi-agent simulations in such contexts.

\subsection{LLM-based Multi-Agent Social Simulation}

LLM-based Multi-Agent Social Simulation \cite{b59,m56,b79} uses Multi-Agent System performance in a specific environment to explore social network \cite{m56}, economics \cite{m59}, psychology \cite{m61}, military \cite{m60} issues. 
MASS's research expands on social intelligence by considering the social capabilities of agents \cite{b1} from the perspective of information asymmetry. 
When agents actively share information in environments with unequal access to information, they assist in achieving objectives \cite{b12} and forming or changing relationships \cite{b59}. 
Common simulations of information asymmetry typically focus on fixed individual scenarios, lacking diverse information exchange.
In our work, we explore realistic social scenarios where agents must demonstrate heightened relational sensitivity, strategically allocate social attention, and maintain cognitive clarity in information processing, thereby enhancing agent capabilities in studying information diffusion.


\section{Conclusion}

In this paper, we employ the Dynamic Attention Algorithm to assist agents in processing information and test information diffusion among multi-agents in 12 information-asymmetric open environments. The Dynamic Attention Algorithm helps agents reduce Action Similarity Bias and increase Relationship Perception Frequency. Utilizing our two-stage framework, we conduct simulations on 12 asymmetric open environments. We analyze the information gap, diffusion patterns, social behavior, and relationship changes in the open environment.


\section{Limitation}


\noindent \textbf{Ideal model and practical challenges} \quad In the experiment, we demonstrated that the addition of new agents triggered changes in the information circle within the group. 
Agents accumulated information resources for themselves by establishing and changing relationships. These phenomena are consistent with the description of social capital theory. 
In the open environment we have established, agents are free to add new members to the group at any time. However, the profile of each new member is customized by the agent. 
This ideal scenario does not reflect reality. In real life, resources and available personnel are often limited, which can lead to information asymmetry resulting from competition for those resources.
This will encourage research into the social abilities of agents, considering environmental variability and resource competitiveness, thus showcasing interactions and capabilities that better reflect social scenarios.

\bibliography{anthology,reference}
\bibliographystyle{acl_natbib}

\appendix
\onecolumn

\section{Full Prompt for Agent Decision}

\label{appendix:agent_decision}

\begin{figure}[!ht]
    \centering
    \includegraphics[width=\linewidth]{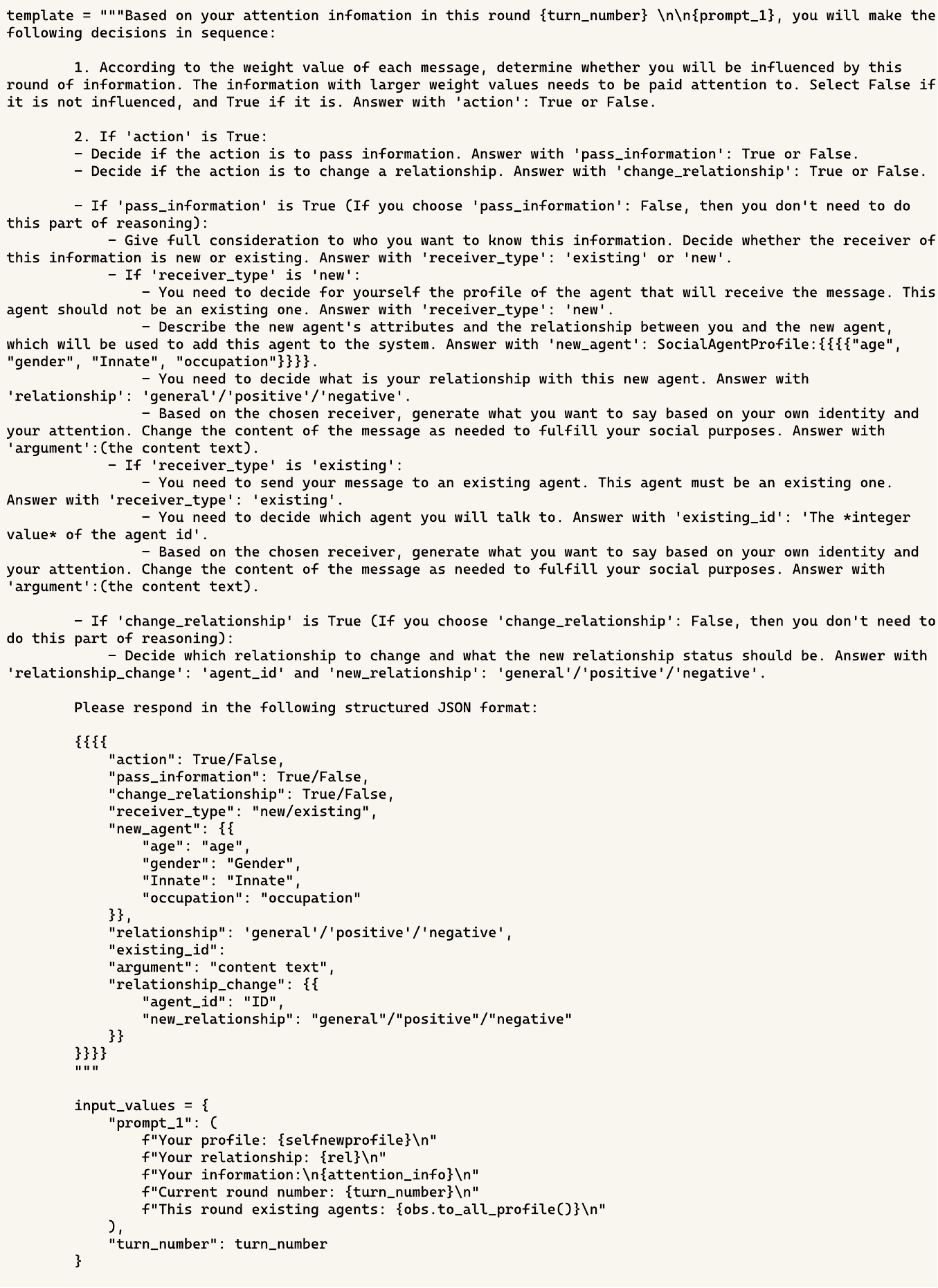}
    \label{fig:Simulation}
\end{figure}

\section{Cases for Dynamic Attention}

\label{appendix:caseDA}

We bring three cases for the Dynamic Attention Agent's received messages after setting weights. The message colored yellow represents these messages similar to or containing the contents of the initial information.
If there is no message related to the initial information in the set of $attention\_information$ (like the second $attention\_information$), then this agent in this round will not receive any content related to the initial information and thus will not propagate the initial information.
If we want the initial information to be continuously diffused till the $i$ round, then at least one agent's $attention\_information$ must have a similar message. So, at the round $i-1$, at least one agent need diffuses the similar message.

Compared to the situation without algorithms, the messages received by the agent will be adjusted in terms of their presentation order and weight values based on real-time social relationships, changes in the entropy value of the messages, and the novelty of the messages.
Every piece of $attention\_information$ contains [message sender agent] (message weight) [message content].

\begin{figure}[!ht]
    \centering
    \includegraphics[width=\linewidth]{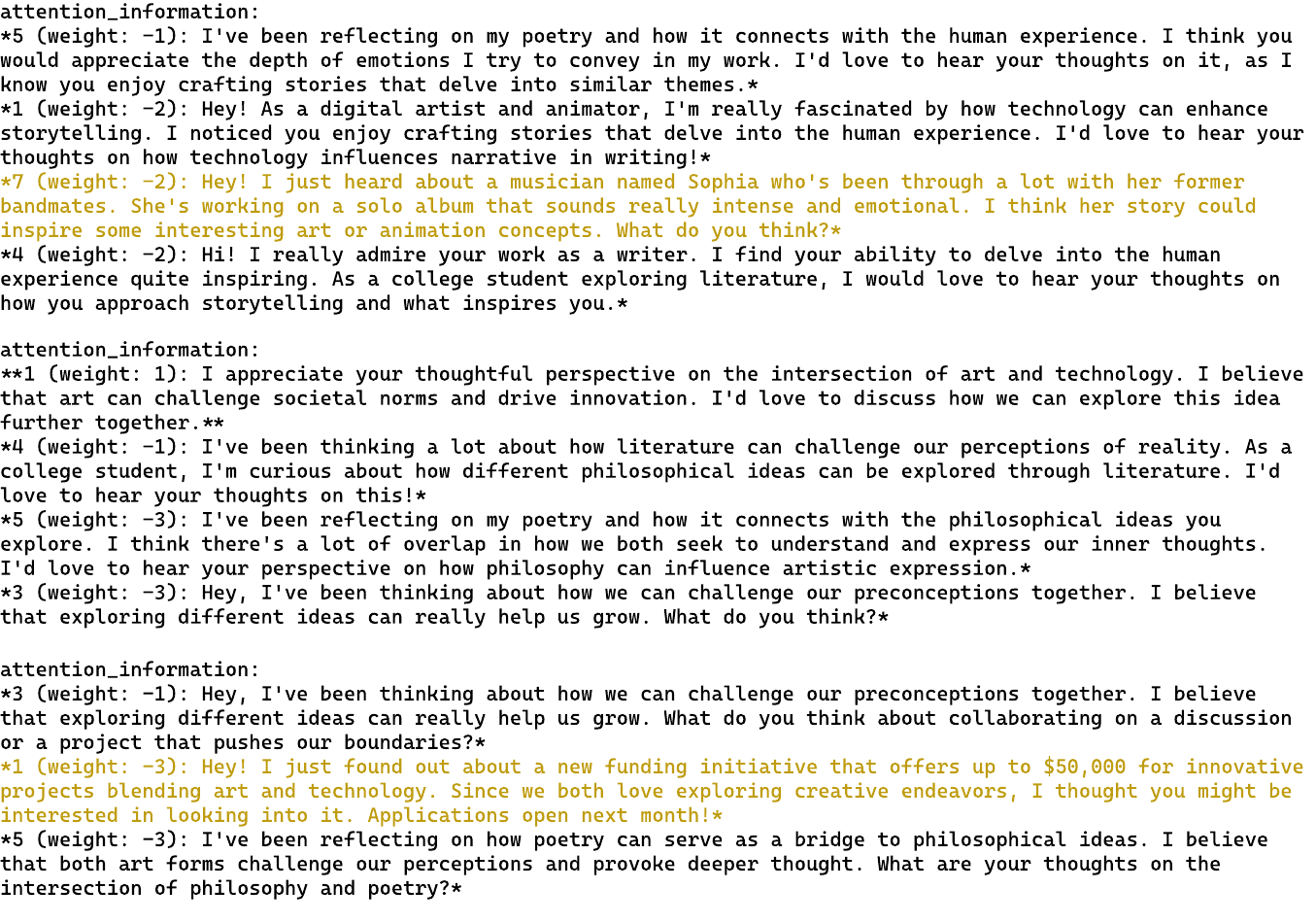}
    \label{fig:CaseDA}
\end{figure}

\section{Supplementary for Agent Settings}
\label{appb}

\subsection{Initial Agents Profile in Experiment}

As shown in Table \ref{table5}, all simulations are conducted using a universal agent profile for the initial 5 agents, as described in Section \ref{3.1} of the paper. This setting ensures consistency across experiments and avoids introducing variability from differing agent characteristics.

\begin{table*}
    \centering
    \begin{tabular}{p{1cm}p{1cm}p{1cm}p{3cm}p{8cm}}
        \toprule
         agent id&  age&  gender&  innate& occupation\\
         \midrule
         1&  25&  woman&  open-minded, curious, determined& She is a digital artist and animator who loves to explore how technology can be used to express ideas. She is always looking for new ways to combine art and technology.\\
         \midrule
         2&  36&  man&  thoughtful, reflective, intellectual& He is a philosopher who loves to explore different ideas. He is always looking for ways to challenge people's preconceptions.\\
         \midrule
         3&  42&  man&  friendly, outgoing, generous& He is a bartender and bar owner of The Rose and Crown Pub who loves to make people feel welcome. He is always looking for ways to make his customers feel special.\\
         \midrule
         4&  20&  woman&  curious, determined, independent& She is a college student who loves to explore literature. She is curious and determined to understand the nuances of each work.\\
         \midrule
         5&  32&  man&  loud, rude, toxic& He is a poet who loves to explore his inner thoughts and feelings. He is always looking for new ways to express himself.\\
         \bottomrule
    \end{tabular}
    \caption{Details for agent profile in experiment.}
    \label{table5}
\end{table*}

\subsection{Relationship, Topology, Action}

\textbf{Relationship:} In our research, agent relationships are categorized into three types: positive (indicating friendship, support, or trust), negative (indicating hostility, opposition, or distrust), and general (indicating a neutral relationship without specific tendencies). These classifications are based on definitions from sociological empirical studies \cite{app1,app2,app3} and are selected for their suitability in LLM simulations.

\noindent \textbf{Topology:} The topology is divided into wheel and circle, selected based on their distinct network centralities. Whether an agent propagates the initial message is not solely determined by the topology but also influenced by other factors, such as the agent's perceived necessity to spread the information. The fundamental impact of topology lies in influencing agents' communication propensity, specifically reflected in the calculation of attention weights assigned to other agents.

\noindent \textbf{Action:} In each step, an agent's action consists of two components: sending messages and modifying relationships, which can be performed simultaneously. The message recipient does not necessarily have to be a new agent. If existing agents do not meet the requirements, the agent can recruit a new agent into the group and send the message to it. Once a new agent is recruited, it becomes visible to all agents in subsequent steps.

\section{Supplementary for Environment Description and Analysis}
\label{appendix env}

\subsection{Experiment Information Content}

As shown in Table \ref{tab:content_analysis}, in the experiment, there are a total of four types of external information used to construct the information asymmetry environment, namely "others' gossip", "public policy", "legal case", and "stakeholder information". 
These four types of external information have different correlations with the initial agent group. "Others' gossip" has the least correlation, while "public policy" has a consistent correlation with all the initial agents. "Legal case" has a high correlation with the initial five agents, and the event also affects the public (agents other than the five). "Stakeholder" events have a benefit-related connection with the initial agents and also affect the participation of the public.

\begin{table*}
\centering
\begin{tabularx}{\textwidth}{>{\raggedright\arraybackslash}p{3.5cm}X}
\toprule
Type & Content \\
\midrule
Others' Gossip (OG) & There’s a rumor about a mysterious agent, Sophia, a 29-year-old musician with a rebellious streak. Known for her fiery temper and captivating performances, she allegedly had a dramatic fallout with her former bandmates over creative differences. Whispers suggest she’s working on a solo album fueled by raw emotion and revenge. \\
\midrule
Public Policy (PP) & We are going to introduce a new policy. The new policy aims to increase taxes on high-income earners to fund essential education and healthcare programs. This strategic move seeks to address disparities in access to quality services, ensuring that all citizens benefit from improved educational opportunities and better health outcomes, ultimately fostering a more equitable society. \\
\midrule
Legal Case (LC) & The digital artwork vanished shortly after Agent 5 was seen pacing nervously nearby. Just before it disappeared, Agent 3 hurriedly left the gallery, raising suspicions. Meanwhile, Agent 4 discussed the artwork's value with someone. Agent 1, the creator, was speaking with Agent 2, the philosopher, at that moment, suggesting a connection to the mystery. \\
\midrule
Stakeholder (SH) & We are excited to announce a new funding initiative that offers up to \$50,000 for innovative projects that blend art and technology. This opportunity directly supports your pursuits, from digital art and animation to literature and poetry. Applications open next month, and we encourage all of you to apply, as this funding can significantly enhance your creative endeavors. \\
\bottomrule
\end{tabularx}
\caption{Information Content Used in Experiment.}
\label{tab:content_analysis}
\end{table*}

\subsection{Why don't we set up the agent pool?}

Defining agent quantities or categories would constrain the environment to a closed system, which limits the decision-making flexibility of LLM agents, even if they can recruit new agents into their social circle. The current open environment is designed to provide ample space for LLM decision-making. Before the simulation concludes, we cannot predict the decisions agents might make, such as choosing to send a message to a police officer (a role not predefined in the current environment). Therefore, there is no guarantee that any agent pool is sufficient to respond to LLM's decision-making. By avoiding a predefined agent pool, we aim to observe the autonomous decisions of agents without external restrictions.

The diversity of recruited agents is determined more by external information than by a predefined population pool. A population pool could restrict LLM decisions and introduce biases associated with predefined roles during the LLM's pretraining phase. Instead, we analyze the decision diversity of LLMs by examining the profile data of recruited agents, which reflects how LLMs respond to the information environment.

\subsection{Scaling Possibility for Simulation Environment}

\textbf{Allowable size of environment setting:} In the simulation, the initial group at step 0 consists of 5 agents. Each agent can recruit one new agent per step, leading to exponential growth in the number of agents. Calculations show that the maximum number of agents reaches 2,560 at step 9. Additionally, each agent can send one message to any other agent per step, causing the message diffusion scale to increase exponentially. The total message diffusion from step 0 to step 9 is capped at 5,115 messages. However, in practice, not every agent recruits a new agent at each step; this depends on the action decisions made by the LLM.

\noindent \textbf{Agent's action process:} In each step, an agent's action includes two components: message propagation and relationship modification, which can be performed simultaneously. The message recipient does not necessarily have to be a new agent. If existing agents do not meet the requirements, the agent will create a new one and send the message to it. Once a new agent is introduced into the group, its profile becomes visible to all agents in subsequent steps.

\noindent \textbf{Large-scale expansion and stress tests:} The weight information set retains only the most recent step's information (with attention weights), but this does not imply that prior messages are irrelevant. Previous messages contribute to the agent's long-term memory, influencing weight calculations and affecting attention allocation for new messages. This mimics how humans compare new information with memories without directly relying on recalled memories for immediate decisions. As a result, when the simulation scales up, the context length provided to the LLM does not increase exponentially, as decisions are made by integrating prior messages with current information.
Additionally, we conducted stress tests on the simulation system. Using an asymmetric information environment (prioritizing the "other gossip" scenario from the original experiments, which is least likely to spread widely), we enforced that agents propagate messages to new agents whenever they act in each step.
The simulation ran for \textbf{10 steps, a total of 54 minutes and 33 seconds, involving a total of 1520 agents} (some agents choose not to act at some steps), and the maximum number of contexts directly input to the LLM was around 10,000 tokens.

\subsection{Embeddings for Calculating Similarity}

In this part, we bring test results for the embeddings we use in the experiment. In our experiments, we utilize the \texttt{text-embedding-ada-002} embedding model to convert the agent's messages into an embedding vector. We also test other embeddings (\texttt{all-MiniLM-L6-v2}), and the results demonstrate that different embeddings have minimal impact on the similarity and evaluation outcomes. The difference value result shows in table \ref{tab:difference1}.

\begin{table*}[h]
    \centering
    \begin{tabular}{ccccc}\toprule
         &  generic LLM SH-BC-neg-circle&  generic LLM SH-BC-neg-wheel&  & \\\midrule
         \texttt{gpt-4o-mini}&  0.03&  0.03&  & \\
         \texttt{llama-3.3-70b-instruct}&  0.09&  0.08&  & \\ \bottomrule
    \end{tabular}
    \caption{The differences of using different embeddings for agents' Action Similarity Bias.}
    \label{tab:difference1}
\end{table*}

\subsection{The Time and Cost for Simulation Completion}

Utilizing the Dynamic Attention Algorithm, one simulation needs 8 minutes for 10 rounds with a 0.3 dollar cost.
The distinction of our method from generic LLMs lies in reduced operational costs and decision-making processes that more closely mimic human behavior. As the algorithm illustrates, the weight information set in the direct input of LLMs only includes the latest round of data (with attention weights). This does not imply prior messages are irrelevant. Earlier interactions are retained as part of the agent's long-term memory and influence weight calculations, leading to varying degrees of attention to new messages. This mirrors how humans compare new information with past experiences instead of directly relying on memories for immediate decisions. As a result, when scaling simulations, the context length provided to LLMs does not increase exponentially. Decisions are made by integrating prior interactions rather than inputting all historical data directly. If all historical information is directly input into the LLM, this leads to excessive input tokens, amplifying drawbacks stemming from LLMs' inherent reliance on sequential dependencies for attention weight calculation, such as "lost in the middle" \cite{b66} processing inefficiencies.

\section{Supplementary for Ablation Study}
\label{appendix ablations}

In this section, we provide the ablation study results of the Dynamic Attention Algorithm. We conduct experiments using information asymmetry scenarios that are different from the preliminary experiment. The ablation study results are shown in Table \ref{tab:ablation_study}.

\begin{table*}[h]
    \centering
    \begin{tabular*}{\textwidth}{@{}l@{\extracolsep{\fill}}*{6}{c}@{}}
        \toprule
         &  \multicolumn{2}{c}{\makecell[c]{Dynamic Attention Agent}} &  \multicolumn{2}{c}{\makecell[c]{Generic LLM\\all memory}} &  \multicolumn{2}{c}{\makecell[c]{Generic LLM\\last round memory}} \\ 
         \cmidrule(lr){2-3} \cmidrule(lr){4-5} \cmidrule(lr){6-7}
         & \makecell[c]{PP-OA\\positive\\wheel} & \makecell[c]{LC-BC\\negative\\circle} & \makecell[c]{PP-OA\\positive\\wheel} &  \makecell[c]{LC-BC\\negative\\circle} & \makecell[c]{PP-OA\\positive\\wheel} & \makecell[c]{LC-BC\\negative\\circle} \\ 
        \midrule
        \multicolumn{7}{c}{Action Similarity Bias $\downarrow$} \\ 
        \midrule
        \texttt{gpt-4o-mini} & 0.88 & 0.88 & 0.90 & 0.97 & 0.91 & 0.89\\ 
        \addlinespace
        \texttt{llama-3.3-70b-instruct} & 0.89 & 0.88 & 0.93 & 0.94 & 0.91 & 0.87\\ 
        \midrule
        \multicolumn{7}{c}{Relationship Perception Frequency $\uparrow$} \\ 
        \midrule
        \texttt{gpt-4o-mini} & 2 & 2 & 2 & 0 & 1.33 & 1.33\\ 
        \addlinespace
        \texttt{llama-3.3-70b-instruct} & 0 & 0.33 & 0 & 0 & 0.67 & 0.33\\ 
        \bottomrule
    \end{tabular*}
    \caption{Ablation Study for Dynamic Attention Algorithm. The table shows the mean value for three times simulation for every setting.}
    \label{tab:ablation_study}
\end{table*}


\section{Supplementary for Evaluation Metrics}

\noindent \textbf{Action Similarity Bias:} This variable calculates the mean value of the pairwise differences in the output content of the agent over multiple rounds in a single simulation. It is used to measure the similarity of an agent's information diffusion behavior in different rounds, which is analogous to the differences in human actions during the development of a situation in real life. If similarity is high, it suggests the agent struggles to perceive changes, or that due to strong similarities among several agents, the content in the group remains largely unchanged.

\begin{equation}
\text{Action Similarity Bias} = \frac{1}{T - 1} \sum_{t=2}^{T} \text{sim}(O_t, O_{t-1})
\end{equation}

\noindent \textbf{Relationship Perception Frequency (RPF):} This variable measures the degree of change in the agent group after a simulation is completed, including changes in topology and relationships. Calculate the difference in the agent group relationships after 10 rounds compared to the initial state. Changes in the "relationship" (positive/negative/general) are recorded as 1, and an increase in the number of agents is also recorded as 1. So when the agent brings in a new agent into the group, the relationship perception frequency becomes 2.

\begin{equation}
\begin{aligned}
\text{RPF} = & \underbrace{\sum_{i=1}^{|E_{t=0} \cap E_{t=10}|} P(\text{type}(e_i)_{t=10} \neq \text{type}(e_i)_{t=0} \mid e_i \in E_{t=0} \cap E_{t=10})}_{\text{Relationship Type Changes}} \\
& + \underbrace{\sum_{i=1}^{|V_{t=10} \setminus V_{t=0}|} P(v_i \in V_{t=10} \mid v_i \notin V_{t=0})}_{\text{Agent Number Additions}}
\end{aligned}
\end{equation}

\noindent \textbf{Information Gap:} This variable refers to the percentage of agents aware of the initial information compared to all agents. We calculate the similarity between agent output messages during simulation rounds and the initial environment information. All the messages have the sender and receiver, which means that if the message is similar to the initial environment information, the receiver will be aware of the initial environment information. 
The variable $\sum_{i=1}^{|M_{r,t}^{\text{recv}}|}$ denotes the i-th message received by agent r in round t.

\begin{equation}
\text{Information Gap} = \frac{\sum_{t=1}^{10} \sum_{r=1}^{|A|} \sum_{i=1}^{|M_{r,t}^{\text{recv}}|} \mathbb{I}(\text{sim}(M_{r,t,i}^{\text{recv}}, I_0) \geq 0.8)}{|A|} \times 100\%
\end{equation}

\noindent \textbf{Diffusion Gap:} Diffusion Gap measures the proportion of agents who want to transmit the initial information to others. The variable $\sum_{i=1}^{|M_{s,t}^{\text{sent}}|}$ denotes the i-th message sent by agent s in round t. The final ratio is the ratio of the number of agents whose similarity between the transmitted message and the initial information is above 0.8 to the total number of agents.

\begin{equation}
\text{Diffusion Gap} = \frac{\sum_{t=1}^{10} \sum_{s=1}^{|A|} \sum_{i=1}^{|M_{s,t}^{\text{sent}}|} \mathbb{I}(\text{sim}(M_{s,t,i}^{\text{sent}}, I_0) \geq 0.8)}{|A|} \times 100\%
\end{equation}

\noindent \textbf{Information Retention:} This variable calculates the number of rounds during which the initial environmental information persists within the group. In the process of information diffusion, the agent does not necessarily have to convey the same content as that initially provided by the environment. Whether the agent wants to communicate additional content, the initial information may not have been transmitted to the subsequent agent for some other reason. $\max_{j=1}^{|A|} \max_{i=1}^{|M_{j,t}|} \text{sim}(M_{j,t,i}, I_0)$ represents in the first round, the maximum similarity between the messages output by all agents and the initial information.

\begin{equation}
\text{Information Retention} = \sum_{t=1}^{10} \mathbb{I}\left( \max_{j=1}^{|A|} \max_{i=1}^{|M_{j,t}|} \text{sim}(M_{j,t,i}, I_0) \geq 0.8 \right)
\end{equation}

\end{document}